\documentclass{article}
\usepackage{spconf,amsmath,graphicx}
\usepackage{amssymb}
\usepackage{comment}
\usepackage{booktabs} 
\usepackage{threeparttable}
\usepackage{xcolor}

\begin{document}
% Title.

\title{NeuroSpex: Neuro-Guided Speaker Extraction with Cross-Modal Fusion}

\name{Dashanka De Silva$^1$, Siqi Cai$^3$, Saurav Pahuja$^1$, Tanja Schultz$^2$, Haizhou Li$^{4,1,3}$}
\address{
\begin{tabular}{c}
$^1$Machine Listening Lab (MLL), University of Bremen, Germany \\
$^2$Cognitive Systems Lab (CSL), University of Bremen, Germany \\
$^3$Department of Electrical and Computer Engineering, National University of Singapore, Singapore \\
$^4$SDS, SRIBD, The Chinese University of Hong Kong, Shenzhen, China
\thanks{This work is supported by the Deutsche Forschungsgemeinschaft (DFG, German Research Foundation) under Germany's Excellence Strategy (University Allowance, EXC 2077, University of Bremen, Germany). This work is also supported by the National Natural Science Foundation of China (Grant No. 62271432) and the Internal Project of Shenzhen Research Institute of Big Data (Grant No. T00120220002).}
\end{tabular}
}

\maketitle

\begin{abstract}
In the study of auditory attention, it has been revealed that there exists a robust correlation between attended speech and elicited neural responses, measurable through electroencephalography (EEG). Therefore, it is possible to use the attention information available within EEG signals to guide the extraction of the target speaker in a cocktail party computationally. In this paper, we present a neuro-guided speaker extraction model, i.e. NeuroSpex,  using the EEG response of the listener as the sole auxiliary reference cue to extract attended speech from monaural speech mixtures. We propose a novel EEG signal encoder that captures the attention information. Additionally, we propose a cross-attention (CA) mechanism to enhance the speech feature representations, generating a speaker extraction mask. Experimental results on a publicly available dataset demonstrate that our proposed model outperforms two baseline models across various evaluation metrics.
\end{abstract}


\section{Introduction}
\label{intro}
Humans inherently can focus on a specific audio source amidst multiple competing sources. This remarkable ability is referred to as selective auditory attention (SAA) in cocktail party scenarios \cite{cite_1}. It was found that two regions of the auditory cortex manage such selective auditory attention. They allow us to separate and enhance the interested voices. However, when we suffer from hearing loss or other forms of impairments, we find it hard to distinguish voices~\cite{cite_3} because our ears are missing some frequencies. Despite much progress, existing algorithms still struggle to isolate and enhance the targeted speech amidst background noise effectively \cite{cite_4}. 

Studies in neuroscience have revealed a strong correlation between the attended speech and the neural response it elicits \cite{cite_6}, paving the way for auditory attention detection (AAD) from neural activities. Such neural responses include Electrocorticography (ECoG) \cite{cite_6}, Magnetoencephalography (MEG) \cite{cite_8}, and EEG \cite{cite_7}. Among them, EEG provides a non-invasive, comparably easy-to-wear, and affordable choice for SAA tasks, especially for neuro-steered hearing aids \cite{cite_12}.

Speech separation and speaker extraction emulate human SAA to solve cocktail party problems. Speech separation algorithms are designed to separate a speech mixture into individual speech streams \cite{cite_19, cite_20}. Conversely, speaker extraction~\cite{cite_21, cite_23} algorithms extract a target speaker's voice associated with an auxiliary reference cue. This cue serves as a distinct marker indicating the specific speech signal of interest to be isolated, typically providing identifiable information about the target or attended speech. Various auxiliary reference cues have been explored to guide speaker extraction systems. Among them, pre-enrolled speech signal \cite{cite_21} from the interested speaker is a commonly used reference signal. Moreover, inspired by human attention mechanisms, visual reference cues derived from video streams capturing hand and body gestures \cite{cite_25}, lip movements \cite{cite_26}, and direction information \cite{cite_27} have been studied. Furthermore, there's a growing interest in multimodal approaches that integrate audio-visual cues \cite{cite_23}. However, in real-world scenarios, it is not feasible to always access pre-enrolled speech from numerous speakers, and visually tracking the target speaker is often impractical.

In this paper, we seek to use the elicited EEG signal as the sole reference cue as it provides feedback from the human brain regarding attended speech. We hypothesize that the neural response can effectively inform a speaker extraction system about the content of the attended speech in real-time. Therefore, we propose a novel end-to-end speaker extraction model at the utterance level in this work, which utilizes the EEG signal temporally aligned with speech as the auxiliary reference cue. We train the model in a trial-independent setting on a public dataset and compare it with several baseline models on multiple evaluation metrics.

This paper is organized as follows. Section 2 summarizes the related work; Section 3 describes the architecture of our proposed model; Section 4 outlines the experimental setup, including the dataset and baseline models. We report our results in Section 5 and conclude the study in Section 6.

\section{Related work}
\label{related work}

Recent advancements in EEG-based AAD and speech separation have opened a path to utilize EEG signals for separate attended speech in a multi-source speech mixture environment. The mixture is first separated into multiple single-speech signals and then the signal with the highest correlation with the EEG signal is selected as the attended speech \cite{cite_28}. However, these methods are limited by the requirement of the number of speakers to be known in advance and the high computational consumption during separation.

In the studies of AAD, a clean speech signal is often compared with the EEG signal to find their correlation. It was shown that \cite{cite_12} the performance of the AAD systems improves as the decision window size increases. However, such clean single-speech signals are not always available in real-world scenarios. This calls for the study on how to exploit the information of the attended speech from within the EEG signals. There have been studies on reconstructing the attended speech envelope \cite{cite_28} from EEG signals. This allows us to establish a relationship between the speech stimulus and its neural responses.  

Multiple end-to-end time domain studies have been proposed addressing different aspects. The Brain-Informed Speech Separation (BISS) \cite{cite_31} uses the reconstructed attended speech envelope from the EEG signal as the reference cue for speaker extraction. Inspired by BISS, the Brain-Enhanced Speech Denoiser (BESD) \cite{cite_32} and follow-up work U-shaped BESD (UBESD) \cite{cite_33} followed a dual-module approach based on Temporal Convolutional Neural network (TCN) to model EEG signals with speech together to denoise the speech mixture to obtain the attended speech. The Brain Assistend Speech Enhancement Network (BASEN) \cite{cite_34} used a TCN and Convolutional multi-layer Cross-Attention (CMCA) module for feature extraction and fusion, respectively. As a follow-up work, sparsity-driver BASEN \cite{cite_35} has proposed two EEG channel selection methods during speech enhancement: residual Gumbel selection \cite{cite_36} and convolutional regularization selection. 

Most recently, the neuro-steered speaker extraction (NeuroHeed) \cite{cite_37} was introduced, which consists of a self-attention (SA) \cite{cite_39} based EEG encoder that generates the reference cue. The speaker extraction task is performed in either an online or offline manner. The online system adopted auto-regressive feedback from previously extracted speech. Following up, the NeuroHeed+ \cite{cite_38} was introduced. It added an auxiliary AAD module to NeuroHeed to reduce the speaker confusion error. 

Despite much progress, it has not been explored in the prior work how both spatial and temporal information from EEG signals can be exploited to build knowledge-rich clues. Moreover, previous research has not studied how to fuse EEG and speech embeddings to make use of the temporal correlation and complementary information of two types of signals during speech mask generation for speaker extraction. We are motivated to explore the above unused potentials for neuro-guided speaker extraction.

% -------------------------- Model Architecture ------------------------
\begin{figure*}[htbp]
\label{NGSEnet_model_architecture}
  \centering
  \begin{minipage}{\textwidth} % Use \textwidth to span over both columns
    \centering
    \includegraphics[width=\linewidth]{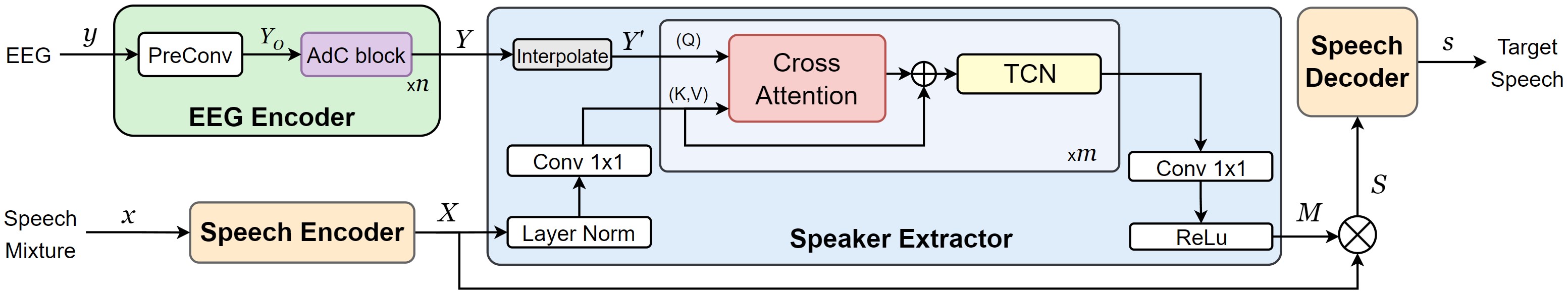} % old file name: NeuroSpex_archi.jpg
    \vspace{-0.5cm}
    \caption{The block diagram of NeuroSpex with all the prominent components and connections. NeuroSpex has $m$ cascaded blocks of cross-attention (CA) and TCN in the Speaker Extractor, and $n$ cascaded AdC blocks in the EEG encoder. {NeuroSpex takes a speech mixture $x$ and EEG signal $y$ as input, generates target speech $s$. Here, $X$ and $S$ represent utterance-level embeddings for speech mixture and target speech, respectively. $Y$ represents the reference signal, where \( Y_{0} \) and \( Y^{\prime} \) denote the output of the pre-convolution (\textit{preConv}) layer and the interpolated reference signal, respectively. $M$ represents the generated mask.} $\oplus$ and $\otimes$ refer to the residual connection and the element-wise multiplication, respectively.}
  \end{minipage}
\end{figure*}

% -------------------------- Cross-Attention and AdC block ------------------------
\begin{figure}[htbp]
\label{cross_attention_and_AdC_block}
    \centering
    \includegraphics[width=0.4\textwidth]{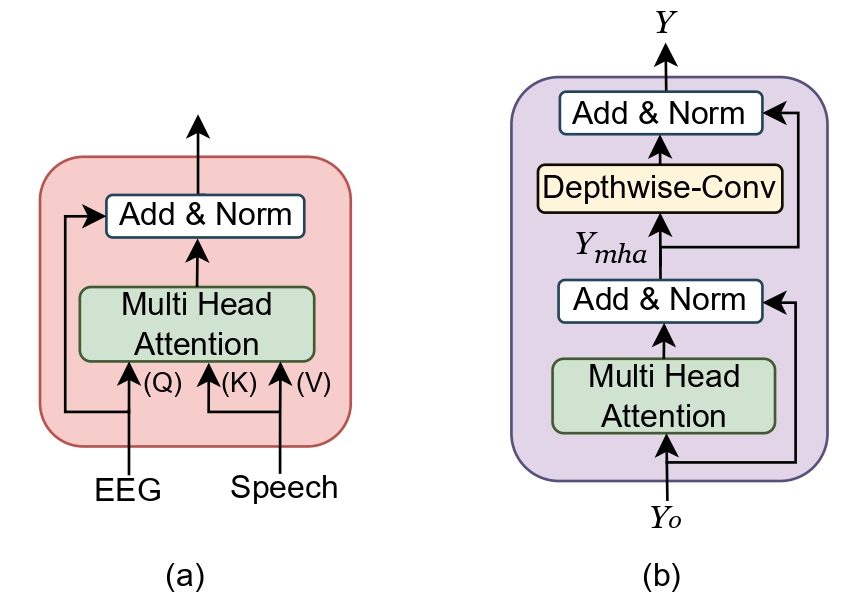}
   \vspace{-0.35cm}
    \caption{(a) CA block to fuse EEG and speech mixture embeddings in the speaker extractor. (b) AdC block from the EEG encoder including multi-head attention and depth-wise convolutions.}
\end{figure}
\vspace{-0.1cm}

\vspace{0.1cm}
\section{Methodology}
\label{methodology }

We propose a neuro-guided speaker extraction algorithm, an end-to-end EEG-based speaker extraction model or NeuroSpex, comprising four components: speech encoder, EEG encoder, speaker extractor, and speech decoder as depicted in Fig.~1. This model is built based on ConvTasNet \cite{cite_19}, consisting of a speech encoder, separator, and decoder. Our model also has an EEG encoder module to guide the extractor by providing information on the attended speech. Given the mixture of attended and interfering speech in the time domain, our model aims to extract the attended speech using elicited neural response as the sole reference cue. 

\subsection{Speech Encoder}

The speech encoder transforms the {single-channel input mixture signal segment $x \in\mathbb{R}^{T_{s}}$ into a sequence of utterance-based embeddings $X \in\mathbb{R}^{N_{x} \times T_{x}}$}, similar to the frequency analysis with short-time Fourier transform. This is achieved by applying time domain speech encoding \cite{cite_19}, that is a 1D Convolutional ($Conv1D$) layer followed by Rectified Linear units (ReLu) as the activation function:
\begin{equation}
    {X = ReLu(Conv1D(x,1,N,K)) \in\mathbb{R}^{N_{x} \times T_{x}}}
\end{equation}

Here, the input channel size is 1 for the single-channel speech utterance with a sequence length $T_{s}$, which is a 4-second segment sampled at 8 kHz (as discussed in Section 4.1). The output $X$ is a sequence of $T_{x}$ embeddings, each with $N_{x}$ dimensions, which are 3200 and 256, respectively. The kernel size $K$ of the Conv1D is set to $20$, while stride and padding are set to $K/2$. 

\subsection{EEG Encoder}
The proposed EEG encoder seeks to exploit complementary information of the attended speech contained in the multi-channel EEG signals and develops the EEG  embeddings or the reference signal to guide the speaker extractor. The EEG encoder takes a 4-second segment of a 64-channel EEG signal $y$, sampled at 128 Hz (detailed in Section 4.1), as input and generates \( Y \in \mathbb{R}^{N_{y} \times T_{y}} \), where \( T_{y} = 512 \) and \( N_{y} = 64 \). This output \( Y \) serves as the reference signal for the speaker extractor. 
    
The EEG encoder has a pre-convolutional layer followed by a series of multi-head attention and depth convolution blocks that serve as AdC blocks as illustrated in Fig. \ref{cross_attention_and_AdC_block} (b). Specifically, the pre-convolutional layer ($preConv$) acts as a preliminary feature extractor, which extracts important features for subsequent layers from EEG signal segment {$y$}. Then we have $n$ number of sequentially layered stack of AdC blocks, where each Adc block consists of a multi-head attention \cite{cite_39} and 1D depthwise-convolutional layer \cite{cite_47}. We employ multi-head attention due to its ability to effectively capture temporal dependencies from sequential data as applied in previous SAA research \cite{cite_40}. Furthermore, motivated by recent work \cite{cite_41} to encode EEG signals, we use depth-wise convolutions to capture frequency-specific spatial information from channels. Both multi-head attention ($mHA$) and depth-convolutional ($dConv$) layers have residual connections followed by layer normalization ($LN$). That is,
\begin{equation}
\label{eeg_enc_eq_1}
    {Y_{0} = preConv(y) \in\mathbb{R}^{N_{y} \times T_{y}}}
\end{equation}
\begin{equation}
\label{eeg_enc_eq_2}
    {Y_{i} = AdC_{i}(Y_{i-1}) \in\mathbb{R}^{N_{y} \times T_{y}}}
\end{equation}
where $Y_{0}$ is the output of the $preConv$ layer that is fed to the first AdC block ($AdC_1$). $Y_{i}$ is the output of $i^{th}$ AdC block where $i \in \left\{1,...,n \right\}$. Each AdC block is designed as follows:
\begin{equation}
{Y_{mha} = LN_{i}^{mHA}\left( mHA_{i}(Y_{i-1},N,H) + Y_{i-1} \right) \in\mathbb{R}^{N_{y} \times T_{y}}}
\end{equation}
\begin{equation}
{Y_{i} = LN_{i}^{dConv}\left( dConv_{i}(Y_{mha},N,K) + Y_{mha} \right) \in\mathbb{R}^{N_{y} \times T_{y}}}
\end{equation}
where $Y_{mha}$ is the output of the  multi-head attention and $N$ is the channel size that is $64$ for all cases. The number of heads of the attention layer $H$ and the kernel size of the depth-convolutional layer $K$ are set to $2$ and $10$, respectively. {The EEG encoder outputs $Y$ from the final AdC block.}

\subsection{Speaker Extractor}

The speaker extractor generates the estimation mask to separate the attended speech from the mixture background. The masked speech $S$ is created by element-wise multiplication of speech mixture embeddings $X$ and the generated mask $M$ as shown in Equation~(\ref{speech_extractor_eq}). {The extracted speech has an embedding sequence length $T_{x}$ of dimension $N_{x}$, the same as the input speech mixture embedding $X$.}
\begin{equation}
\label{speech_extractor_eq}
    {S = X \otimes M \in\mathbb{R}^{N_{x} \times T_{x}}}
\end{equation}
As shown in Fig. 1, we propose a speaker extractor based on the ConvTasNet backbone with TCNs \cite{cite_19} to increase the receptive field, and CA modules to fuse speech mixture and EEG embeddings as shown in Fig. 2 (a). We adopt cross-modal attention~\cite{cite_39} to integrate information from both the auditory stimuli (speech mixture) and brain responses (EEG), mirroring the brain's natural process of speech mixture perception during speech comprehension. 

The cross-attention (CA) mechanism is designed to combine and enhance insights from the key-value given the query~\cite{cite_16}. Before fusing EEG and speech mixture embeddings, the EEG {embedding} (i.e., the reference signal $Y$) is linearly interpolated from length $T_{y}$ to $T_{x}$ (512 to 3200) to match the embedding sequence length required by the CA block, thereby generating $Y^{\prime} \in\mathbb{R}^{N_{y} \times T_{x}}$. Subsequently, the key and value inputs to the CA block are the speech mixture embeddings $X$, while the query input is the interpolated reference signal $Y^{\prime}$. The output of the CA block $X_{ca}$ is computed as shown in Equation~\ref{CA_eq}, where Q, K, and V are query, key, and value, respectively. The $X_{ca}$ also has the same dimensionality of $X$ and is then fed to the TCN block with a residual connection.
\begin{equation}
\label{CA_eq}
    {X_{ca} = CA(K=X, V=X, Q=Y^{\prime}) \in\mathbb{R}^{N_{x} \times T_{x}}}
\end{equation}
As shown in Fig. 1, we repeat CA and TCN pairs 4 times $(m = 4)$ in the speaker extractor. 

\subsection{Speech Decoder}

The speech decoder reconstructs time-domain single-channel speech waveform $s$ from masked speech embeddings $S$. It performs an inverse operation of the speech encoder by passing $S$ through a linear layer ($linear$) and an Overlap-and-Add ($OvlpAd$) function to restore the audio signal from speech representation. That is,
\begin{equation}
\label{speech_decoder_eq}
    {s = OvlpAd(linear(S,N,L), L\mathbin{/}2) \in\mathbb{R}^{T_{s}}}
\end{equation}
{where $T_{s}$ denotes the length of the output speech utterance.} The linear layer has input size of $N$ and output size of $L$, which are 64 and 20, respectively. The Overlap-and-Add operation also has a $L/2$ frame shift.

\subsection{Loss Function}

We train the proposed model end-to-end using the scale-invariant signal-to-distortion ratio (SI-SDR) \cite{cite_42} between the reference attended speech signal $s_t$ and the extracted speech signal $s_e$. SI-SDR loss is typically used in time-domain speaker extraction tasks \cite{cite_19, cite_37}. The extracted speech is scaled to remove changes that occurred during the reconstruction and helps maintain stability during the training phase. It is calculated with dB and a higher SI-SDR indicates better speech quality. Therefore, a negative SI-SDR is used as the loss function to train the model. The SI-SDR is defined as follows:
\begin{equation}
    \mathcal{L}_{SI-SDR}(s_t,s_e) = -10\log_{10}{(\frac{ \parallel \frac{s_e^{T}s_t}{\parallel s_t \parallel^{2}}s_t \parallel^{2} }{\parallel s_e- \frac{s_e^{T}s_t}{\parallel s_t \parallel^{2}}s_t \parallel^{2}})}
\end{equation}

\vspace{0.2cm}
\section{Experimental Setup}
\label{experimental setup}
\subsection{Dataset}

We used the publicly available KULeuven (KUL) \cite{cite_43} dataset, which includes EEG signals from 16 normal-hearing subjects, collected using the BioSemi ActiveTwo system with 64 channels at a sampling rate of 8,192 Hz. The dataset comprises 20 trials per subject, though our experiments used only the first 8 to avoid repetition in attending to the same speech stimulus. Each trial involves subjects listening to simultaneous speech recordings from two male speakers narrating Dutch short stories, delivered dichotically via plugged-in earphones. Subjects focused on the speech from one speaker while ignoring the other. The stimuli, generated from four stories, were balanced for loudness and presented in randomized speaker direction and identity. The speech signals are sampled at $8$ kHz and mixture signals are created by mixing attended and unattended speech signals at $0$ dB to have the same power for both speakers.

EEG recordings were referenced to the average of all electrodes, band-pass filtered between 1 and 32 Hz, and downsampled to 128 Hz. This frequency range was selected to align with AAD protocols. Data normalization was performed trial-by-trial to standardize mean and variance across the dataset. In total, the dataset comprises 128 trials amounting to 12.8 hours of speech-EEG parallel data.

We followed a speaker-dependent trial-independent training approach to train all configurations of the proposed NeuroSpex and baseline models. Therefore, we divided the data into three sets: train, test, and validation. For the test set, we randomly selected 1 trial from each subject, thus having 16 trials for all subjects. Since each trial has a pair of attended and unattended audios, we made sure no same pair is chosen more than twice from all subjects. Similarly, from the rest of the data, we randomly selected 4 trials to form the validation set. The remaining trials are used for the training set.

Each trial was cut into 4-second segments with a hop length of 1 second for all sets based on experimental observations and computational resources. This segmentation approach yielded 5,712 segments for the test set, 1,428 segments for the validation set, and 38,556 segments for the training set. Specifically, the test set comprised data from 16 subjects, with each subject contributing 357 segments, totaling 5,712 segments, each lasting 4 seconds. We train the model on the training data and evaluate its performance on the validation set at the end of each epoch to monitor progress and adjust hyper-parameters accordingly. After training, the test set is used to obtain the final evaluation metrics to report the performance.

\vspace{-0.2cm}
\subsection{Evaluation Metrics}
\label{eval_metrics}
We employ three evaluation metrics and an extension of one of them. SI-SDR \cite{cite_42} and SI-SDR improvement (SI-SDRi) which quantifies the improvement in the quality of the extracted speech signal compared to the mixture signal. Perceptual Evaluation of Speech Quality (PESQ) \cite{cite_44} assesses the quality of extracted speech by comparing it to the clean single speech signal, providing a score indicating intelligibility and naturalness. Short-Term Objective Intelligibility (STOI) \cite{cite_45} evaluates the intelligibility of extracted speech by comparing it to the clean single speech signal, providing a measure of speech clarity and understandability. All evaluation metrics are higher the better.

\vspace{-0.2cm}
\subsection{Baseline Models}
We benchmark NeuroSpex against several baseline models. The baselines are justified as follows.

Firstly, we use blind speech separation with permutation invariant training BSS (PIT) \cite{cite_46} model based on DPRNN \cite{cite_20}. {This model blindly separates a multi-talker speech mixture into multiple single-talker streams. It then selects the target speech by comparing the best permutation with the target speech based on the SI-SDR metric, without relying on a reference cue to guide the separation. This method is optimized to separate speech streams in a discriminative manner. It represents the full speaker extraction potential when we have perfect neural decoding. Thus, it is seen as the upper bound of NeuroSpex in terms of speaker extraction performance.

Secondly, two versions of NeuroHeed are considered, which operate with a speaker extractor. They are DPRNN \cite{cite_20} and ConvTasNet \cite{cite_19} based on recurrent neural networks and TCNs, respectively. NeuroHeed is the current state-of-the-art for neural response-based speaker extraction. NeuroHeed with ConvTasNet is similar to NeruroSpex in terms of model architecture, while NeuroHead with DPRNN has fewer parameters than NeuroSpex. They make two relevant baselines for our benchmarking.

Thirdly, we use the BASEN model which also adopts TCN and CMCA EEG-Speech processing in terms of model architecture. Each model has been used in its vanilla form and trained in the same conditions as NeuroSpex. BASEN is being chosen as a baseline also because it represents the state of the art before NeuroHeed, and it also adopts TCN and CA in its architecture.

\vspace{-0.2cm}
\subsection{Model Training}

All implementations utilize the PyTorch framework with distributed training and data distributed sampler techniques across 2 Nvidia RTX A6000 GPUs. To ensure reproducibility, we used random seeds to generate consistent results. The models are trained end-to-end using the Adam optimizer with an initial learning rate of 0.0001. A learning rate scheduler with a decay factor of 0.5 when the best validation loss does not improve within 5 consecutive epochs and early stopping is applied when the best validation loss does not improve within the last 25 epochs. Training is conducted for around 100 epochs or until stable training with a batch size of 16. 

\begin{table}[t] % [htbp]
\centering
\caption{
Ablation study to gain insights into the contribution of different model components to overall performance utilizing multiple metrics evaluated on {the validation set}.
}
\vspace{0.2cm}
\begin{tabular}{lccccc}
\hline\hline
EEG & Fusion & SI-SDR & SI-SDRi & PESQ  & STOI  \\
encoder &&&&& \\
\hline
direct & direct & 7.167 & 9.583 & 1.706  & 0.714   \\
$\text{AdC}_{1}$ & direct & 12.819 & 15.981 & 2.088  & 0.813   \\
 SA & CA & 13.894 & 17.063 & 2.243  & 0.835  \\
 Conv & CA & 14.911 & 18.380 & 2.438  & 0.856   \\
$\text{AdC}_{1}$ & CA & 16.605 & 19.775 & 2.464 & 0.872  \\
\hline\hline
\end{tabular}
{\begin{tablenotes}
    \small
    \item[1] $\text{AdC}_{1}$ : EEG encoder with a single AdC block.
\end{tablenotes}}
\vspace{-0.1cm}
\label{ablation_study}
\end{table}
%-----------------------------------------------------------------------------

%-------------------------------- AdC modules Table ------------------------------------------
\begin{table}[htbp]
\centering
\caption{Impact of the number of AdC blocks on the performance of our proposed NeuroSpex, evaluated using multiple metrics on the test set}
\vspace{0.2cm}
\resizebox{0.46\textwidth}{!}{
\begin{tabular}{ccccc}
\hline\hline
AdC blocks  & SI-SDR & SI-SDRi & PESQ  & STOI  \\
\hline
 1 & 16.756 & 19.873 & 2.476 & 0.882  \\
 2 & 17.332 & 20.552 & 2.589  & 0.889   \\
 4 & 17.155 & 20.375 & 2.559  & 0.886   \\
 6 & 17.489 & 20.709 & 2.592  & 0.893   \\
 8 & 17.120 & 20.330 & 2.531  & 0.886   \\
\hline\hline
\end{tabular}}
\label{NGSE_model_configuerations}
\end{table}
%-----------------------------------------------------------------------------

\section{Results}
\label{results}

In this section, we discuss the empirical evaluations of our study by comparing our model with baseline models, exploring different model configurations, and conducting an ablation study. Performance on the trial-independent test and validation sets is reported using the evaluation metrics outlined in Section~\ref{eval_metrics}. The results are summarized in tables and illustrated with SI-SDRi violin plots. Statistical significance was assessed using paired \textit{t}-tests. 

\subsection{Ablation Study}

%-------------------------------- Model Comparison Table ------------------------------------------
\begin{table*}[ht]
\label{Model_comparison_table}
\centering
\caption{Performance evaluation of the proposed NeuroSpex against the baseline models, with different speaker extractors and EEG encoders in terms of SI-SDR and SI-SDRi for extracted speech quality, and PESQ and STOI for speech intelligibility. The results are reported for the test set with the number of parameters in millions.}
\vspace{0.2cm}
\resizebox{0.9\textwidth}{!}{
\begin{tabular}{lcccccccc}
\hline\hline
Model & Extractor & EEG encoder & Fusion  & SI-SDR & SI-SDRi & PESQ & STOI & \# Parameters \\
\hline
BSS (PIT) & DPRNN & - & - & 21.351 & 24.047 & 2.873 & 0.956 & 2.64M \\
NeuroHeed & ConvTasNet & SA & direct & 13.926 & 17.146 & 2.441  & 0.831 & 9.26M \\ 
NeuroHeed & DPRNN & SA & direct & 8.249 & 11.472 & 2.240  & 0.774  & 2.88M \\ 
BASEN & TCN & CMCA & CMCA & 6.241 & 8.534 & 1.285  & 0.347  & 663K \\
\textcolor{black}{$\text{NeuroSpex}_{1}$} & ConvTasNet & $\text{AdC}_{1}$ & CA & 16.756 & 19.873 & 2.476 & 0.882  & \textcolor{black}{5.00M}\\  
$\text{NeuroSpex}_{6}$ & ConvTasNet & $\text{AdC}_{6}$ & CA & 17.489 & 20.709 & 2.592 & 0.893  & 5.09M \\  
\hline\hline
\end{tabular}
}
\begin{tablenotes}
    \small
    \item[1] $\text{AdC}{1}$: EEG encoder with one AdC block; $\text{AdC}{6}$: EEG encoder with six AdC blocks.
    \item[2] $\text{NeuroSpex}_{6}$: Proposed NeuroSpex model with the EEG encoder with 6 AdC blocks (best performing).
\end{tablenotes}
\end{table*}
%-----------------------------------------------------------------------------

We conducted the ablation study to understand the effect of the contributing components. Table \ref{ablation_study} shows the performance of models, evaluated using multiple metrics on the \textcolor{black}{validation set}, for different EEG encoders and feature fusion methods including direct EEG signal input and direct fusion. Note that we only used a single AdC block ($AdC_{1}$) in the EEG encoder. The results mainly show that introducing CA to speech-EEG embeddings fusion increases the performance on every evaluation metric. Furthermore, AdC block also improves the performance over SA used in NeuroHeed and Convolution-based EEG encoders.

%---------------------------- Violin Plot: Subject Comparison -----------------------------
\begin{figure}[ht] 
\label{NeuroSpex_per_subject_compare_graph}
  \centering
    \begin{minipage}[b]{1.0\linewidth}
		\centering
		\centerline{\includegraphics[width=8.5cm]{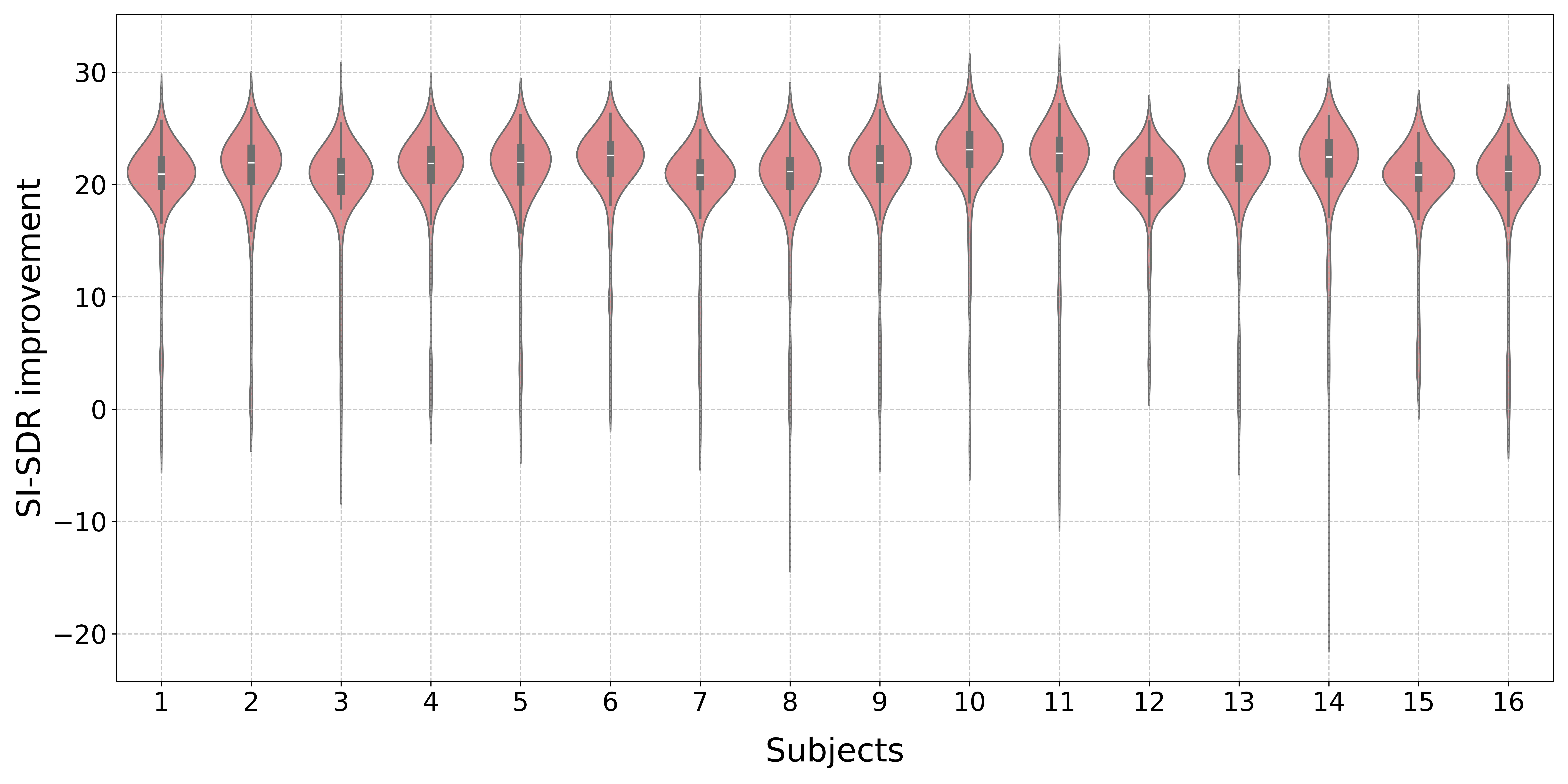}}
  \vspace{-0.35cm}
    \caption{Violin plots of extracted speech SI-SDR improvement for each subject from the test set. Consistency in speech output is observed across subjects.}
	\end{minipage}
 \vspace{-0.4cm}
\end{figure}
%-----------------------------------------------------------------------------

%--------------------------------- Violin Plot: Model comparison ----------------------------------
\begin{figure}[ht] 
\label{baseline_model_compare_graph}
  \centering
    \begin{minipage}[b]{1.0\linewidth}
		\centering	\centerline{\includegraphics[width=8.5cm]{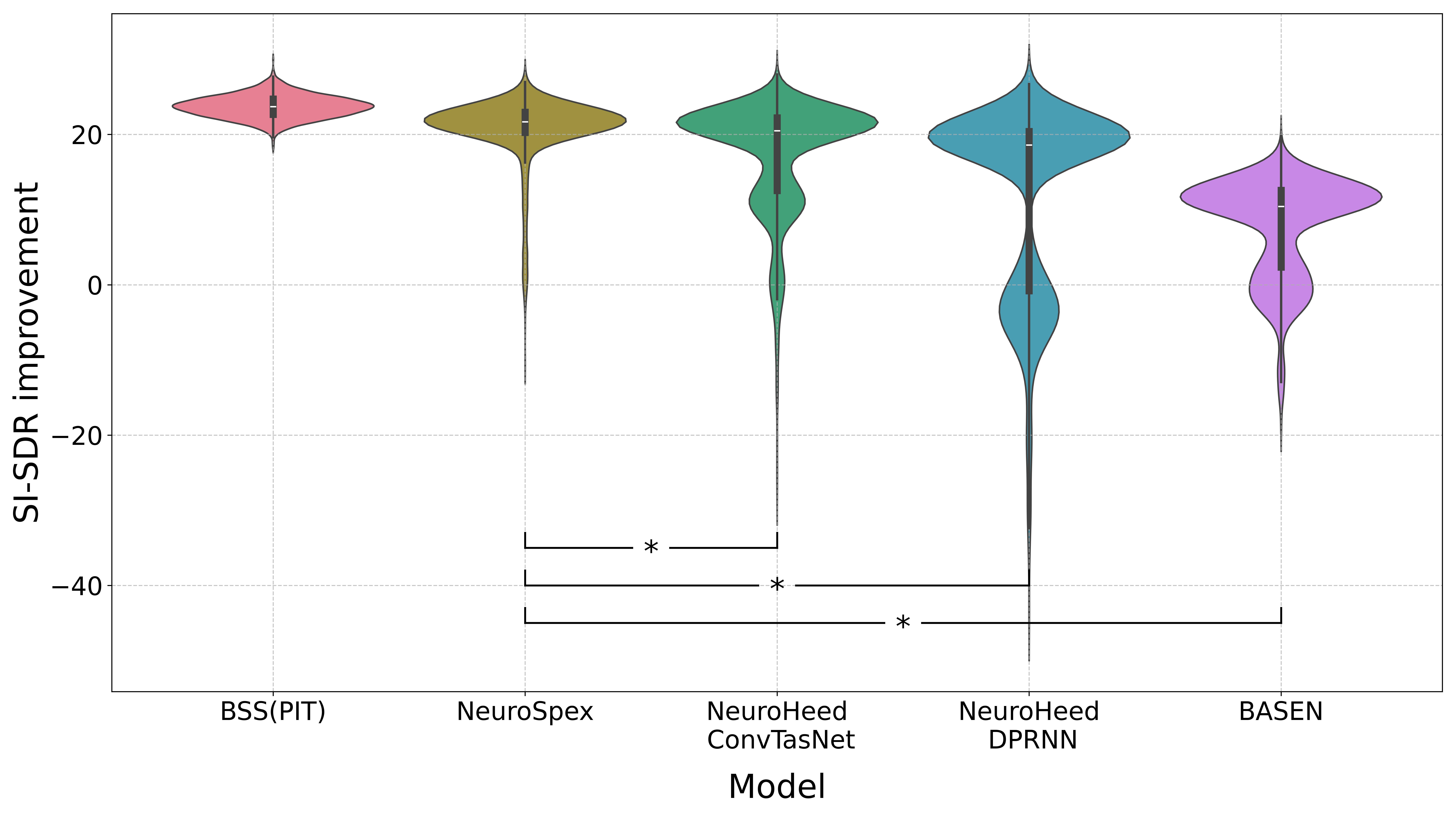}}
  \vspace{-0.35cm}
    \caption{Violin plots of extracted speech SI-SDR improvement from the test set for NeuroSpex and 4 baseline models to compare output performances.  Here, $\ast$ represents the statistical significance of the comparison (p \textless 0.001, paired \textit{t}-test)}
	\end{minipage}
\vspace{-0.4cm}
\end{figure}

%-----------------------------------------------------------------------------

\subsection{Analysis of AdC Blocks}
We evaluated the effect of the number of AdC blocks in the EEG encoder to find the appropriate number. Table~\ref{NGSE_model_configuerations} reports the performance for 5 progressive number of AdC blocks. For all evaluation metrics, 6 AdC blocks return the best performance. This shows that the number of AdC blocks increases the performance from 1 to 6 blocks with smaller margins. Hence, we present our best-proposed model with 6 AdC blocks. Furthermore, Fig. 3 depicts the violin plots of SI-SDRi for each of the 16 subjects in the test set for our best-proposed model. All subjects show consistent and centered distribution with smaller variances except for subjects 8, 11, and 14.

\subsection{Comparison with Baseline Models}
We compare our best model (with 6 AdC blocks) with multiple baseline models as shown in Table 3 including the main architectural difference between models and evaluation metrics. The violin plots in Fig. 4 summarize the SI-SDRi for all trials in the test set for all models. The BSS (PIT) shows the best results with the least variance as it performs direct speech separation and sets an upper bound for the comparison. Our proposed model outperforms all baselines significantly (\textit{p} $<$ 0.001) except BSS (PIT) on all evaluation metrics showing that it performs the speaker extraction with better signal quality perpetual quality, and intelligibility. Furthermore, the proposed NeuroSpex model contains fewer parameters compared to the ConvTasNet-based NeuroHeed model but more than DPRNN counterparts due to the TCN-based speaker extractor. According to violin plots, the proposed model has a better-centered distribution around a higher median and less variance compared to both NeuroHeed and BASEN baselines.

\section{Conclusion}
\label{conclusion}
In this study, we propose an end-to-end speaker extraction model operating in the time domain. The model utilizes neural responses as a reference cue to extract attended speech in a cocktail party scenario. The novel contributions of this work include multi-head attention and depth-wise convolution-based EEG encoding, and CA-based EEG-speech embeddings fusion, which seek to enhance the overall quality of speaker extraction. The results demonstrate significant improvements in extraction performance over several competitive baseline models. Thus, we show that our proposed model effectively extracts EEG embeddings correlated with attended speech and achieves superior speech-EEG feature fusion to generate the speaker extraction mask, hence validating our hypothesis. For future research, we recommend exploring speaker-specific information during extraction and conducting subject-independent studies to enhance generalizability and realism.

% -------------------------------------------------------------------------
\bibliographystyle{IEEEbib}
\bibliography{refs}

\end{document}